\documentclass[manuscript]{acmart}
\usepackage{comment}
\usepackage{graphicx}
\usepackage{moreverb,url}
\usepackage{float}
\usepackage{url}
\usepackage{multirow}
\usepackage{colortbl}
\usepackage[acronym]{glossaries}

\usepackage{todonotes}

\usepackage{hyperref}

\usepackage[super]{nth}

\usepackage{listings}
\usepackage{listings-golang} 

\AtBeginDocument{%
  \providecommand\BibTeX{{%
    \normalfont B\kern-0.5em{\scshape i\kern-0.25em b}\kern-0.8em\TeX}}}

\setcopyright{acmcopyright}
\copyrightyear{2024}
\acmYear{2024}
\acmDOI{XXXXXXX.XXXXXXX}

\begin{document}

\title[Preserving Automotive Heritage: A Blockchain-Based Solution for Secure Documentation of Classic Cars Restoration]{Preserving Automotive Heritage: A Blockchain-Based Solution
\\for Secure Documentation of Classic Cars Restoration}

\author{José Murta}
\affiliation{%
  \institution{NOVA University Lisbon - School of Science and Technology, NOVA LINCS}
  \city{Caparica}
  \country{Portugal}
}
\email{j.murta@campus.fct.unl.pt}

\author{Vasco Amaral}
\orcid{0000-0003-3791-5151}
\affiliation{%
  \institution{NOVA University Lisbon - School of Science and Technology, NOVA LINCS}
  \city{Caparica}
  \country{Portugal}
}
\email{vma@fct.unl.pt}

\author{Fernando Brito e Abreu}
\orcid{0000-0002-9086-4122}
\affiliation{%
  \institution{Iscte – Instituto Universitário de Lisboa, ISTAR-IUL}
  \city{Lisbon}
  \country{Portugal}
}
\email{fba@iscte-iul.pt}

\begin{abstract}

    Classic automobiles are an important part of the automotive industry and represent the historical and technological achievements of certain eras. However, to be considered masterpieces, they must be maintained in pristine condition or restored according to strict guidelines applied by expert services. Therefore, all data about restoration processes and other relevant information about these vehicles must be rigorously documented to ensure their verifiability and immutability. Here, we report on our ongoing research to adequately provide such capabilities to the classic car ecosystem.
    
    Using a design science research approach, we have developed a blockchain-based solution using Hyperledger Fabric that facilitates the proper recording of classic car information, restoration procedures applied, and all related documentation by ensuring that this data is immutable and trustworthy while promoting collaboration between interested parties. This solution was validated and received positive feedback from various entities in the classic car sector. The enhanced and secured documentation is expected to contribute to the digital transformation of the classic car sector, promote authenticity and trustworthiness, and ultimately increase the market value of classic cars.
    
\end{abstract}



\received{March 2024}
\received[revised]{DD Month YYYY}
\received[accepted]{DD Month YYYY}

\maketitle


\typeout{NT FILE acronyms.tex}%


%

\newacronym{NOVA}{NOVA}{NOVA University Lisbon}

\newacronym{FCT}{FCT}{NOVA School of Science and Technology}

\newacronym{DI}{DI}{Department of Computer Science}

\newacronym{ACP}{ACP}{\href{https://www.acp.pt/classicos}{Automóvel Club de Portugal}}

\newacronym{ASE}{ASE}{Automated Software Engineering}

\newacronym{FIVA}{FIVA}{\href{https://fiva.org/en/fiva-home/}{Fédération Internationale des Véhicules Anciens}}

\newacronym{IoT}{IoT}{Internet of Things}

\newacronym{DSR}{DSR}{Design Science Research}

\newacronym{RR}{RR}{Rapid Review}

\newacronym{SLR}{SLR}{Systematic Literature Review}

\newacronym{P2P}{P2P}{Peer-to-Peer}

\newacronym{DLT}{DLT}{Distributed Ledger Technology}

\newacronym{PoW}{PoW}{Proof of Work}

\newacronym{PoS}{PoS}{Proof of Stake}

\newacronym{PBFT}{PBFT}{Practical Byzantine Fault Tolerance}

\newacronym{MSP}{MSP}{Membership Service Provider}

\newacronym{GUI}{GUI}{Graphical User Interface}

\newacronym{BPMN}{BPMN}{Business Process Model and Notation}

\newacronym{DMN}{DMN}{Decision Model and Notation}

\newacronym{CRM}{CRM}{Customer Relationship Management}

\newacronym{ML}{ML}{Machine Learning}

\newacronym{CA}{CA}{Certificate Authority}

\newacronym{VIN}{VIN}{Vehicle Identification Number}

\newacronym{SRE}{SRE}{Software Requirements Engineering}

\newacronym{PoA}{PoA}{Proof of Authority}

\newacronym{PICOC}{PICOC}{Population, Intervention, Comparison, Outcomes and Context}

\newacronym{IPFS}{IPFS}{InterPlanetary File System}

\newacronym{CFT}{CFT}{Crash Fault Tolerance}

\newacronym{NFT}{NFT}{Non-Fungible Token}

\newacronym{API}{API}{Application Programming Interface}

\newacronym{INCD}{INCD}{Infraestrutura Nacional de Computação Distribuída}

\newacronym{SSH}{SSH}{Secure Shell}

\newacronym{CID}{CID}{Content Identifier}

\newacronym{REST}{REST}{Representational State Transfer}

\newacronym{HTTP}{HTTP}{Hypertext Transfer Protocol}

\newacronym{HTTPS}{HTTPS}{Hypertext Transfer Protocol Secure}

\newacronym{SMTP}{SMTP}{Simple Mail Transfer Protocol}

\newacronym{SDK}{SDK}{Software Development Kit}

\newacronym{JWT}{JWT}{\gls{JSON} Web Token}

\newacronym{JSON}{JSON}{JavaScript Object Notation}

\newacronym{TPS}{TPS}{Transactions Per Second}

\newacronym{SUS}{SUS}{System Usability Scale}

\newacronym{GDPR}{GDPR}{General Data Protection Regulation}



\section{Introduction}
\label{sec:introduction}

Classic automobiles are much more than old cars. The materials used in their original construction, the craftsmanship that went into building them, and the historical significance they may hold make them true works of art. Like other pieces of artwork, classic vehicles tend to be collectable items and can be sold for millions of euros. 

To consider vintage car masterpieces and consequently associate them with a large price tag, they need to be equipped with meticulous records, and most of them must be restored following strict guidelines and expert restoration services. These guidelines can be proposed by both national entities, such as \gls{ACP}, and international ones, like \gls{FIVA}, which published a guide providing practical advice on restoration and maintenance of historic vehicles called the 'Charter of Turin Handbook' \cite{charter_of_turin}.

The restoration process can significantly impact the automobile's price and overall quality, so the restoration's history, with all the activities performed during each process and other important information, must be properly recorded in tamper-proof, immutable documentation. This documentation can be used for certification or evaluation by classic car owners, workshops, and certification institutions.

The lack of trust and transparency and the need to provide rigorous information in the classic car sector is the problem we are trying to tackle with our work, offering a solution using blockchain technology to document immutable, transparent, and tamper-proof information about restoration procedures, with all related pieces of evidence, like media files or any other type of vehicle documents, together with an appropriate access control mechanism, capable of effectively handling diverse scenarios within the classic car industry, achieving an immutable and trustworthy portfolio of the car's history. This solution allows classic car owners to prove the authenticity of their vehicles and restorers to validate their work while streamlining the certification process for authorities. This enhances vehicle value, recognition, and confidence in the classic car sector.

Our proposed work aligns with an ongoing project \cite{RodrigoGomes2023, PedroMoura2023} that aims to digitally transform a classic car shop by developing a platform dedicated to tracking the progress of classic car restorations. This web platform enables customers to monitor the progress of their vehicles and access documentation generated throughout the restoration process. The system uses machine learning algorithms and sensors to automatically identify the type of restoration activities conducted on the vehicles, creating Business Process Model and Notation (BPMN) and Decision Model and Notation (DMN) models that represent the restoration processes supporting the guidelines from \textit{FIVA}'s Charter of Turin \cite{charter_of_turin}, this way naming this platform as \textit{Charter of Turin Monitor}. Please refer to the cited papers for additional information about this project.

The \gls{DSR} methodology has been followed in this work. This methodology is a systematic, iterative and scientific outcome-based approach that aims to acquire the necessary knowledge to create new artifacts to solve a well-defined problem in a specific research area. It provides specific guidelines for evaluation within research projects \cite{dsrNovoArtigo, design_SR2}. Thus, in this paper, we: (i) identify and motivate the problem, (ii) define the objectives of a solution to this problem, (iii) describe the design and development of the corresponding artifact, (iv) present how we demonstrated its functionalities to obtain constructive feedback, and (v) how we evaluated the effectiveness, usability, and performance of this artifact. This paper also aims to partially fulfil the last step of the \gls{DSR} methodology, called "communication", which stands for the need to share with the community the findings, results, and lessons learned.

The rest of this paper is organized as follows: in section \ref{sec:background}, some insight into important concepts to work being developed is provided; next, in section \ref{sec:slr}, we present the result of a systematic \gls{RR} of published literature with similar objectives, both in the automotive industry and in other sectors; next, in section \ref{sec:swRequirements} we present and discuss the proposed solution, including its stakeholders and the software architecture; in section \ref{sec:implementation} it is detailed the implementation decisions; in section \ref{sec:v_and_v}, we describe how we evaluated performance and usability and interpret the corresponding results; finally in section \ref{sec:conclusions}, we review the main achievements and discuss potential future work.

\section{Background} \label{sec:background}

\textbf{Blockchain} technology is often described as an exceptional solution for the longstanding challenge of establishing trust among humans \cite{Bashir2020}. It is a decentralized and distributed digital ledger that allows the recording and verification of transactions without a central authority. It is a distributed system that stores, manages, and shares data among all participants in the network, maintaining a continuously growing list of transaction data records, which are cryptographically secured to prevent tampering and revision. 

Blockchain technology was first proposed in a white paper published by Satoshi Nakamoto in 2008 and was introduced as the underlying technology for the Bitcoin cryptocurrency \cite{bitcoin_satoshi}. This paper detailed a peer-to-peer (P2P) online payment system that, by establishing an immutable, decentralized public ledger that records all transactions and forbids double-spending, eliminating the need for banks.

Understanding blockchain requires familiarity with its key technologies. Though not new, their combination makes blockchain unique. These include:
\begin{itemize}
    \item \textbf{Cryptography} is essential for almost every application that runs on the Internet. It establishes secure digital identities for each network participant and verifies transactions through cryptographic hashing and digital signatures, ensuring authentication, non-repudiation, and data integrity;
    
    \item \textbf{Distributed Ledger Technology} makes it possible to maintain information securely and decentralized by logging asset transactions on a shared database;
    
    \item \textbf{Consensus Protocols} determine how a network agrees on the state of a ledger and validates transactions. After an update on the ledger, all the nodes in the network must have a single and accurate copy of the ledger, but this is only possible after the nodes reach a consensus on which version of the ledger is correct. The most popular consensus protocols are the \textit{Proof-of-Work (PoW)} which was introduced to the blockchain technology with Bitcoin's creation, \textit{Proof of Stake (PoS)}, which consists of a viable alternative to PoW and which the \textit{Ethereum} blockchain recently adopted, and \textit{Crash Fault Tolerance (CFT)} and \textit{Practical Byzantine Fault Tolerance (PBFT)}, consensus protocols widely used in permissioned blockchains.
\end{itemize}

Together, these technologies and their development make blockchain a technology with unique and distinctive characteristics that have gained interest among several sectors all around the globe. These characteristics are decentralization, immutability, transparency, and integrity in trustless environments. Another technology that has significantly contributed to the widespread acceptance of blockchain is \textbf{smart contracts}. Characterized as code deployed and operational within the blockchain, smart contracts can automate the execution of functions when predefined terms and conditions of an agreement are met \cite{blockchainAndSmartContracts}.

%
Blockchain systems can be categorized into three types: public, private, or consortium, depending on their use, requirements, and access permission levels.

\textbf{Public} or \textbf{permissionless} blockchains are typically open source and available to anyone worldwide without restrictions. Network participants can read, write, and participate in the consensus process. This blockchain type provides economic incentives for users to participate in the consensus process to attract more participants to join the network. Bitcoin blockchain is the first and most notorious implementation of a public blockchain.

A \textbf{private} blockchain is a digital ledger that provides proprietary networks where a central organization can grant access to the potential participants. Within an organization, a specified group of users holds centralized write permissions, whereas read permissions can be restricted or available to the public. Industries specifically design this kind of blockchain for their internal and commercial due to its effectiveness and auditability. Ethereum and \textit{Hyperledger Fabric} are the most commonly used platforms to deploy and maintain private blockchains.

A \textbf{consortium} or \textbf{hybrid} blockchain is derived from the two previously mentioned types. As the name suggests, a group of organizations collaborates to maintain the network in this blockchain category. In contrast to writing permissions, which are only given to verified and acknowledged participants representing the associated organizations, read permissions can be public or restricted. This type of ledger is best suited for inter-organizational collaboration. As in private blockchains, the most popular and well-suited platforms for consortium blockchains are Ethereum and Hyperledger Fabric.

\section{Related Work}
\label{sec:slr}

A \gls{RR}, also called Rapid Literature Review, is a research methodology to identify, examine, and evaluate all the available findings, evidence, and analyses of other authors on a specific research question or question or topic area \cite{Tricco2015}. In our work, a state-of-the-art rapid literature review was performed to guarantee detailed research on this study's topics, providing the necessary background on the subject by obtaining relevant evidence about the available technologies and limitations in current research. Later in this section, in \autoref{subsub:greyLiterature}, we will explain how this literature review also encompasses elements and guidelines from a Multivocal Literature Review \cite{multivocal}. For the sake of simplicity, this paper presents only the primary steps and key results of the review. 

Defining the research questions is a crucial step in planning any literature review to help later the researchers identify primary studies that assess these questions, apply the criteria of inclusion, and extract and analyze the necessary data to answer them. 

The following \textbf{research questions} were formulated to obtain the most adequate answers possible through our review:
\begin{itemize}
    \item \textbf{RQ1}: How is blockchain technology used for documentation and evidence purposes?
    \item \textbf{RQ2}: How is documentation stored in trustworthy systems that are not prone to tampering?
    \item \textbf{RQ3}: How is blockchain technology used in the vehicle industry?
    \item \textbf{RQ4}: Is blockchain being explored for documentation purposes in the vehicle industry, especially in the classic car domain?
    \item \textbf{RQ5}: Which network types and blockchain platforms are used in evidence documentation?
\end{itemize}

With the research questions formulated, we analysed and performed the actual steps of the \gls{RR} methodology, defining a clear review protocol and formulating the search strings to be executed in our literature databases. The results from these searches as well as the literature found using a snowboarding approach \cite{Wohlin2014} resulted in a total of 68 articles gathered. 

All the studies were analyzed to determine whether or not they were relevant to our research. Initial inclusion criteria were applied, followed by an assessment of the quality of each study's content. After these phases, a total of 9 studies found relevant to our topic and field of research were accepted.

\subsection{Generic Primary Studies} \label{solutionsResearched}

Several solutions in the literature present a novel implementation using blockchain technology to facilitate, store, and generate trustworthy documentation. Although it is applied in different industries, such as healthcare and supply chain, it has similar requirements.

Zhong et al. \cite{Zhong2022} proposed a framework that uses blockchain technology to manage the impact of hospitals during the COVID-19 pandemic by tracking infection control measures. The framework ensures an immutable, tamper-proof record of events and authenticates records. Hyperledger Fabric is the selected platform for implementing the blockchain network. The blockchain is a consortium blockchain, where different organizations with different permission levels interact. The framework is supported by off-chain cloud data storage to store larger data files that are unsuitable for blockchain storage.

Authors in \cite{Ismail2019} propose a lightweight blockchain architecture for healthcare data management. A PBFT protocol is used to achieve consensus on the ledger. As patient data is sensitive, the authors opted for a permissioned blockchain network that limits network participation to authorized members. However, this architecture does not support smart contracts, which is a significant limitation compared to Hyperledger Fabric or the Ethereum platform. Though promising, this solution's potential is constrained due to the absence of smart contract functionality.

Zhang et al. \cite{Zhang2020} propose a system architecture based on blockchain technology to track data of all the operations involved in the entire grain supply chain to guarantee food quality. The authors designed a multimode storage mechanism to improve blockchain storage efficiency. The blockchain network, developed with Hyperledger Fabric, ensures the information is traceable and credible. Moreover, each blockchain node includes an off-chain database to address storage limitations for large or diverse data types or data not requiring consensus.

In \cite{Ferdousi2020}, authors introduced a smart contract-enabled supply chain framework on a permissioned blockchain network. It's originally tailored for the US beef cattle industry and adaptable to other supply chains with minimal adjustments. The system records all supply chain operations in the blockchain for immutable, confidential, and tamper-proof data. Built on the Ethereum platform, the solution utilizes a P2P network, where each node hosts both the blockchain and requires a local database service for raw farm data storage.

Although in completely different sectors and with completely distinct use cases, the technologies presented in all these studies can inspire the work that we are developing.

\subsection{Blockchain-based Vehicle Documentation Solutions} \label{sec:vehiclesSolutions}

To be more precise about the applications of blockchain-based solutions to the automotive industry, few relevant studies are available in the literature that utilize blockchain technology to register information about vehicles. A significant amount of the studies are oversimplified proofs of concept that do not have any kind of real application. Furthermore, some of the published literature is being applied in several areas of the automotive industry beyond the scope of our research, such as in enhancing communication in autonomous vehicles. Those other applications are not reviewed here.

\subsubsection{Grey Literature} \label{subsub:greyLiterature}

As mentioned at the beginning of this chapter, this literature review can also be categorized as a Multivocal Literature Review since not only formally published scientific literature was analyzed but also included grey literature, which involves the review of websites, blogs, white papers, and proprietary solutions \cite{multivocal}. By adopting this multivocal approach, our review bridges the gap between pure scientific research and industry projects, incorporating cutting-edge solutions applied in real-world applications.

\href{https://vinchain.io/}{\textit{VINchain}} is a blockchain-based solution to transparently and securely record important information about vehicles. The information about a specific vehicle is retrieved using its VIN (Vehicle Identification Number). However, it is necessary to pay to obtain the desired information. This solution does not implement efficient data access management mechanisms since anyone with access to the vehicle's VIN and willing to pay for the associated data will have it. In contrast, even the vehicle owner would have to pay for information about his vehicle. This access fee model raises concerns about data privacy, as certain vehicle information may need to be kept confidential to prevent undesirable actions by malicious participants.

To finish the review of commercial implementations with a similar scope, it is also crucial to present \href{https://themotorchain.com/en/}{\textit{The Motor Chain}}. The Motor Chain is the platform most similar to our solution since it is geared towards the classic vehicles sector. It is advertised as a blockchain-based platform where classic car owners can register their vehicles, and any updates on the documentation performed by the owners or professional entities (e.g., maintenance services, restoration shops, and appraisal entities) are reflected in the vehicle documentation timeline. The Motor Chain claims to use Hyperledger Fabric as its chosen blockchain platform to record all the desired information. However, due to its proprietary nature, there exists no accessible means to verify the actual usage of this platform or the technology. Furthermore, there is still insufficient information accessible on the access management mechanism, and the car owners have to pay to register their vehicles in the system. Another point of consideration lies in the platform's primary focus on a basic timeline of classic car events, lacking adherence to \gls{FIVA}'s Charter of Turin guidelines. To the best of our knowledge, this system lacks the capability to incorporate other types of media evidence besides photos, and information on media storage is unavailable. Despite this, The Motor Chain shares similar goals, reinforcing the need for our work and aiding in identifying challenges.

\subsection{Formal Scientific Literature}

Regarding the scientific studies using blockchain technology to store and secure trustworthy information about vehicles, our focus of interest, we left out those not peer-reviewed, oversimplified, without practical use, or using outdated blockchain platforms or technologies that no longer work due to discontinued components or developer abandonment. Notwithstanding, some relevant research explores blockchain technology in securing immutable vehicle documentation, as follows. 

Y. T. Jiang and H. M. Sun \cite{Jiang2021} proposed a blockchain-based system to create a trusted source of vehicle data for the second-hand vehicle market. The system uses a consortium blockchain network built on the \textit{Ethereum} platform and utilizes the \textit{Proof of Authority} (PoA) consensus protocol. The PoA protocol verifies and combines transactions in blocks through selected credible entities, similar to the Practical Byzantine Fault Tolerance protocol. The system provides access control management, allowing different entities, such as car dealers, repair shops, and government branches, to interact with specific data stored on the ledger. Each vehicle is identified by its unique VIN, and a token provided by the car dealer is required to access its data record. Since the system uses \textit{Ethereum}, recording data on the ledger incurs a transaction fee in the native cryptocurrency \textit{Ether}. 

In \cite{Brousmiche2018V2} and \cite{Brousmiche2018}, from the same authors, a vehicle data and process ledger framework was proposed to facilitate collaboration and secure sharing of vehicle maintenance history among multiple stakeholders over a consortium blockchain. This system aims to furnish a transparent record of a vehicle's history throughout its life cycle, starting from the original automotive manufacturer and ending with the car owner, by documenting all relevant operations, such as repairs, maintenance, and accidents. Access to a vehicle’s data is restricted first to the manufacturer, car owner, and brand official repair shops. This access can be granted to other entities, such as third-party maintenance shops or insurance companies. This solution mixes blockchain technology and classic databases for storage purposes. The authors chose \textit{Quorum} as the blockchain platform for this implementation. Quorum enables control nodes to access the ledger using a whitelist, validates blocks using a specially designed consensus protocol that does not require mining, and removes the cost of executing transactions on the blockchain.

In \cite{Luchoomun2020}, the authors proposed a blockchain-based smart contract application to avoid some of the common challenges of the automotive industry, such as tampering with vehicle information and falsifying mileage. With the proposed solution, the vehicles' data is transparent and immutable while offering the possibility to streamline some processes, leading to increased trust among customers. This solution's blockchain network implementation consists of Hyperledger Fabric. The authors conducted interviews and surveys with automotive industry stakeholders, gathering information and opinions that underscore the need for our research, identifying a lack of trust in sellers, and a general agreement that blockchain technology would significantly improve the sector.

G. Subramanian introduced another relevant solution, and A. S. Thamp \cite{Subramanian2021}. The authors proposed a consortium blockchain solution similar to the ones already presented but focusing on the electric car market, where the major actors are the vehicle manufacturer, charging station, and battery manufacturer. Since electric vehicles and their charging stations have more software support, smart IoT sensors enabled automatic interaction with the blockchain, such as recording vehicle mileage during charging. Ethereum serves as the blockchain platform for this solution, with each transaction incurring an associated cost. 

\autoref{tab:relatedWork} provides a concise overview of the formally reviewed scientific literature within the automotive industry, highlighting the contributions and limitations of each work and offering a comparative analysis.

\begin{table}[]
\centering
\caption{Comparison of formal related work on the automobile context.}
\label{tab:relatedWork}
\resizebox{\textwidth}{!}{%
\begin{tabular}{ccccccccc}
\hline
\rowcolor[HTML]{EFEFEF} 
\textbf{Work} & \textbf{Author(s)}                                                                      & \textbf{Year} & \textbf{Purpose}                                                                          & \textbf{\begin{tabular}[c]{@{}c@{}}Blockchain\\ category\end{tabular}} & \textbf{\begin{tabular}[c]{@{}c@{}}Blockchain\\ platform\end{tabular}} & \textbf{\begin{tabular}[c]{@{}c@{}}Transaction\\ fee\end{tabular}} & \textbf{\begin{tabular}[c]{@{}c@{}}Documentation\\ detail\end{tabular}} & \textbf{\begin{tabular}[c]{@{}c@{}}Applied to \\ Classic's\\ Industry\end{tabular}} \\ \hline
{\cite{Jiang2021}}    & \begin{tabular}[c]{@{}c@{}}Y. T. Jiang and \\ H. M. Sun\end{tabular}                    & 2021          & \begin{tabular}[c]{@{}c@{}}Second-Hand\\ vehicle market\end{tabular}                      & Public                                                                 & Ethereum                                                               & Yes                                                                & Medium                                                                  & No                                                                                  \\ \hline
{\cite{Brousmiche2018, Brousmiche2018V2}}   & \begin{tabular}[c]{@{}c@{}}K. L. Brousmiche \\ et al\end{tabular}                       & 2018          & \begin{tabular}[c]{@{}c@{}}Vehicle's history \\ throughout its \\ life cycle\end{tabular} & Consortium                                                             & Quorum                                                                 & No                                                                 & Medium                                                                  & No                                                                                  \\ \hline
{\cite{Luchoomun2020}}      & \begin{tabular}[c]{@{}c@{}}K. Louchoomun, \\ S. Pudaruth and \\ S. Kishnah\end{tabular} & 2020          & \begin{tabular}[c]{@{}c@{}}Prevent \\ falsification \\ of mileage\end{tabular}            & Consortium                                                             & \begin{tabular}[c]{@{}c@{}}Hyperledger\\ Fabric\end{tabular}           & No                                                                 & Low                                                                     & No                                                                                  \\ \hline
{\cite{Subramanian2021}}      & \begin{tabular}[c]{@{}c@{}}G.Submanian and \\ A.S. Thampy\end{tabular}                  & 2021          & \begin{tabular}[c]{@{}c@{}}Lifecycle of \\ electric/smart \\ vehicles\end{tabular}        & Consortium                                                             & Ethereum                                                               & Yes                                                                & Low                                                                     & No                                                                                  \\ \hline
\end{tabular}%
}
\end{table}

\subsection{Summary of the Research} \label{sec:summaryResearch}

Through our literature review, we identified and understood the technologies needed to tackle similar issues, enabling us to draw important conclusions. Even though certain literature examined was used in various fields, not specifically relevant to the automotive industry, or not formally published due to the multivocal approach of this review, it is vital in understanding cutting-edge technologies and served as a source of inspiration for our work. This process has provided valuable insights into the context and current research related to documentation within the automotive sector. It was also possible to conclude that consortium is the network type preferred to record reliable documentation and that Hyperledger Fabric and Ethereum, or derivations from it, are the favored blockchain platforms.

Some relevant studies, projects, and initiatives specific to the automotive sector aim to preserve vehicle data, documentation, and overall history using blockchain technology. However, none of these have been adequately tailored to meet the specific needs and demands of the classic automobile context. These solutions primarily target the conventional automotive market, and the recorded data is typically plain and straightforward, like maintenance logs. The solution most tailored to the classic vehicle industry is \textit{The Motor Chain}. However, it offers only a relatively limited timeline of events with no minute details. It presents other issues identified previously, such as uncertainty about the authenticity of the information and the entity responsible for its registration. 

Both classic car owners and the certification authorities, along with the restoration procedures, require a higher level of meticulousness, namely a precise and detailed record for every stage of the restoration process. Proper documentation outlining each step of the restoration process, including supporting evidence and the responsible parties, grants authenticity that, as for most works of art, is highly considered for determining the market value. 

\section{System Proposal}
\label{sec:swRequirements}

This section aims to present the conceptualization and solution proposal of this work by presenting the software requirements and stakeholders and delineating the architecture and components of the proposed system.

\subsection{Software Requirements Engineering} \label{sub:sre}

\gls{SRE} is the disciplined application of proven principles, methods, and notations to discover, analyze, and document the services and constraints a system should provide based on the needs of customers and stakeholders \cite{futureSWEngineering}. The \gls{SRE} process should occur before any actual designing, coding, or testing takes place.

We started by defining who is affected and who can affect our system by identifying the distinct stakeholders, who have different roles and requirements and different levels of access and permissions to interact with the system \cite{aurum2005engineering}. 

Although we did not use formal techniques for stakeholder mapping, requirements elicitation drew upon the valuable expertise and experience of industry specialists. The stakeholders identified their priorities and requirements as follows:

\begin{enumerate}
    \item \textbf{Classic Car Owner}: can register themselves as new users in the system; register a new classic car resorting to a certified restoration shop or certification authority; view the documentation of each owned vehicle, which we named \textit{Vehicle Card}; view the chronological update history of the \textit{Vehicle Card,} along with information on the individuals or entities responsible for these modifications; add documentation to its vehicles when desired to add legal documents; grant permissions to other users and entities; finally, can change the ownership of its vehicles' records to another user;
    
    \item \textbf{Certified Restoration Shop Administrator}: The restoration shops will be responsible for most of the restoration and conservation procedures, so they need to have the required access to modify the documentation of the vehicle;
    
    \item \textbf{Certification Body Officer}: Certification regulators of classic cars are responsible for proper vehicle certification after its restoration or repair. For this purpose, they require permission to check vehicle documentation. They will also need permission to mark the documentation as certified by the authority;
    
    \item \textbf{Potential Buyer}: As previously mentioned, possible buyers of a classic vehicle can be granted permission to check the documentation and history of that car, and ultimately, if the buyer decides to buy the car, he gets complete access to the car’s history, while the original owners get his permissions revoked.
\end{enumerate}

\subsection{Architecture} \label{subsec:arch}

\begin{figure}[htbp]
	\centering
	\includegraphics[width=.85\linewidth]{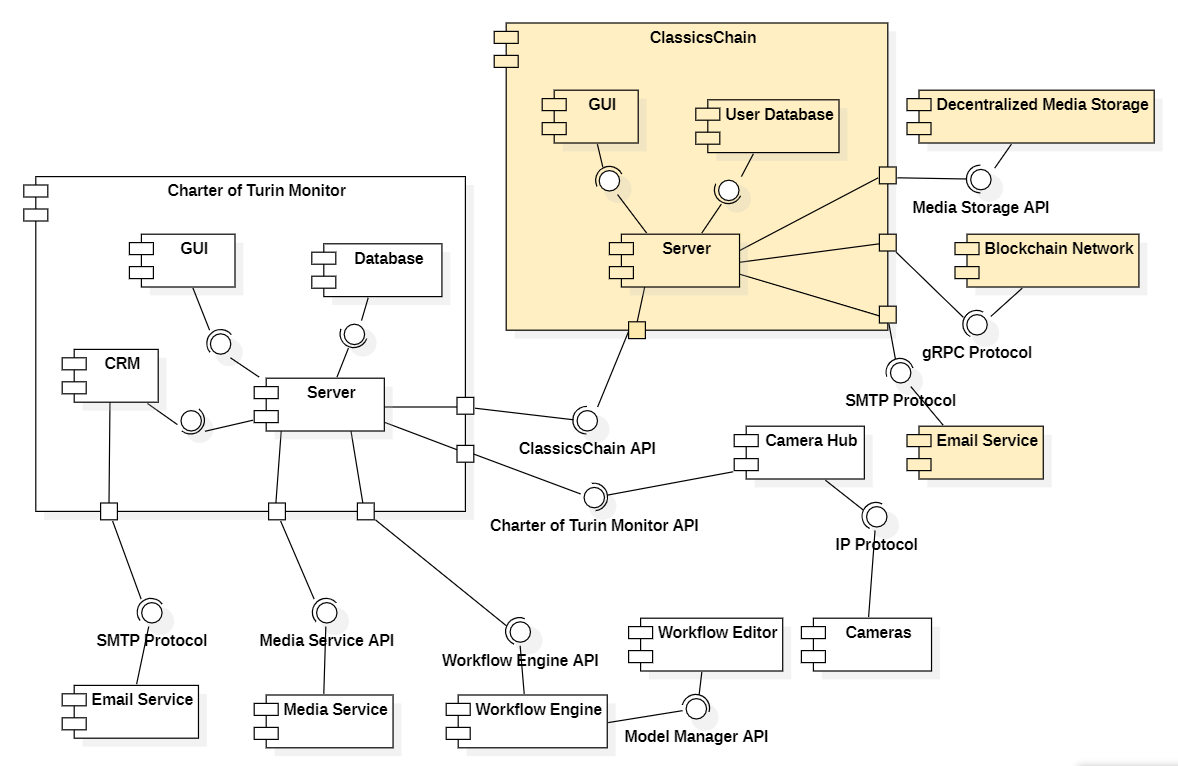}
	\caption{Component Diagram of our system, based on the component diagram in \cite{PedroMoura2023}. The colored components and artifacts identify the work developed in the current research.}
	\label{fig:componentDiagram}
\end{figure}

With the requirements of our project clearly defined, it is appropriate to start explaining the developed system's structure, which relies on these constraints. The component diagram in \autoref{fig:componentDiagram} is an overview of the current system's architecture. It is an upgraded version of the initial component diagram for the system used at the classic restoration shop \cite{PedroMoura2023}, the \textit{Charter of Turin Monitor}. This project involved developing various artifacts, identifiable by their colored components. Although the component diagram depicts them as integrated into the repair shop platform, they can function as an independent system for other entities. 

The developed system was named \textit{ClassicsChain} to highlight the dual focal points of this project: classic cars and blockchain. This solution, its main components and the technologies used are next described:

\begin{enumerate}
    \item \textbf{Blockchain Network}: To securely and immutably preserve data about every detail of all the documentation of restoration processes, such as procedures, materials, tools used, and the hash of other generated pieces of evidence, such as media files. Smart contracts are used to record the data. For this purpose, the chosen technology was \href{https://www.hyperledger.org/use/fabric}{\textbf{\textit{Hyperledger Fabric}}}. Fabric is a consortium-oriented blockchain platform that is highly configurable, presenting increasing scalability and performance. Fabric allows for creating organizations comprising different users for entities connected to the network. It additionally facilitates effective communication among specific network members with ensured data isolation and confidentiality.
    
    \item\textbf{Web Server}: A Linux web server connects users with all other components. It accepts HTTPS requests from client applications and communicates with the blockchain network and the rest of the components. These requests are granted through a middleware \textit{REST API}.
    
    \item\textbf{Decentralized Media Storage}: As our system is based on blockchain, it is also important to store media files associated with classic decentralized cars. However, blockchain is not suitable for high-volume data storage. To address this issue, \textbf{\textit{InterPlanetary File System}} (\href{https://ipfs.tech/}{\textbf{\textit{IPFS}})} is used as off-chain media storage.
    
    \item\textbf{User Database}: To facilitate user registration and, consequently, authentication on the system with typical credentials, besides the registration within the blockchain platform that records the user as an entity of the network, the user is also automatically registered in a user database. This decision was also influenced by some of the constraints imposed by the \gls{GDPR}.
    
    \item\textbf{GUI}: A Graphical User Interface (GUI) is necessary for the web application of this system, allowing users to interact with different functionalities based on their roles. The vehicles' information is displayed as a portfolio with all the relevant data that can function as an Identification Card of the vehicle, which we called \textit{Vehicle Card}. We decided to use \href{https://angular.io/}{\textbf{\textit{Angular}}} for our front-end since the software available at the repair shop where our solution is being integrated uses this technology as its front-end.

    \item\textbf{Email Service}: Lastly, the \textit{Email Service} component is the service that will be used by the \textit{ClassicsChain} system to enable custom emails to be sent automatically, informing users about specific operations that took place.
\end{enumerate}

\begin{figure}[!ht]
    \centering
    \includegraphics[width=1.0\textwidth]{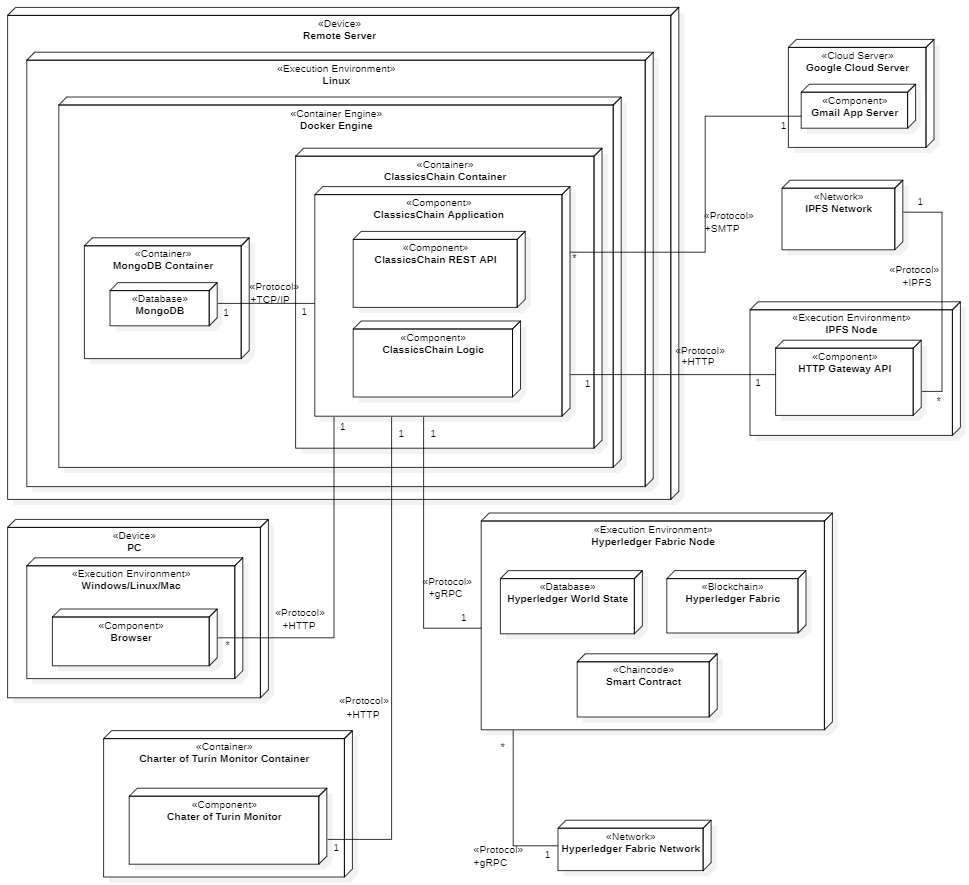}
    \caption{Deployment Diagram of our solution.}
    \label{fig:deploymentDiagram}
\end{figure}

The component diagram of our system's architecture and the technologies employed were combined to generate the deployment diagram in \autoref{fig:deploymentDiagram}.

\section{Implementation} \label{sec:implementation}

This section will delve into key implementation decisions and details of our system, making it uniquely suited for the classic automotive industry. The following subsections will highlight the most significant components of our solution.

\subsection{Blockchain Network}

The blockchain network is the foundational element of our solution, particularly a Hyperledger Fabric Network. As previously described, a Fabric blockchain network is a technical infrastructure that provides ledger services for distributed applications \cite{hyperledgerDocs}. 

\begin{figure}[htbp]
	\centering
	\includegraphics[width=.8\linewidth]{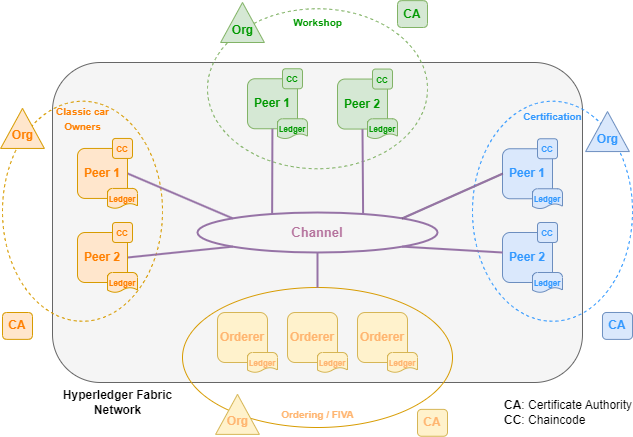}
	\caption{Hyperledger Fabric Network Design Diagram.}
	\label{fig:hlfNetwork}
\end{figure}

Our Hyperledger Fabric network topology is illustrated in figure \ref{fig:hlfNetwork}, with detailed component descriptions provided in the following paragraphs. The network's infrastructure and configuration align with a consortium network. However, this type of implementation is recently being referred to as \textbf{public permissioned}, as a consortium of organizations maintains it. Still, public users affiliated with one of these organizations execute the transactions and read operations.

Our implementation consists of 3 \textbf{peer organizations}: car owners, classic car workshops, and certification entities. Each peer organization comprises 2 peers that maintain the network and execute chaincode (essentially the smart contracts). All peers are anchor peers, meaning they can all be connected through web applications to ensure redundancy, load balancing of requests, and fault tolerance.

It is also mandatory to have an ordering organization. The \textbf{orderer organization} comprises 3 orderer nodes to ensure redundancy and fault tolerance. The orderers are essentially the nodes that perform the network's consensus protocol, utilizing the RAFT protocol, a standard protocol for Fabric networks known for its fault tolerance. Given its impartiality and lack of interest in disrupting the consensus protocol, it would be most beneficial if this organization is maintained by a trusted entity like \gls{FIVA}.

Regardless of being a peer or orderer, each organization has a \gls{CA} responsible for generating authentication certificates for all network components.

Another important decision in our implementation was using only one channel. A channel essentially serves as the communication medium between components, and each channel represents a ledger.

Finally, it is important to note that our network's deployment was not merely an experimental local deployment. It is a production-ready blockchain network managed with a Kubernetes cluster facilitated by a cloud provider for streamlined deployment.

\subsection{Smart Contracts}

Smart contracts, or chaincode in Hyperledger Fabric nomenclature, are also one of the cornerstones of this solution. They control the entire backend, performing functions related to our data models, services, and logic. They handle tasks such as recording information through transactions, reading via queries, and access control. They were written in Golang, and although they can be composed in other languages, Go is the most refined and widely adopted by the community.

Assets are defined as Go structs to represent the distinct objects on the ledger. To encapsulate the mentioned fundamental elements of our system, we specified the following 3 structs or assets in our smart contracts: \textbf{Classic}, which represents the fundamental details of a classic car, such as registration number, model, and VIN number (which acts as the primary key of each classic); \textbf{RestorationStep}, which represents the details of a task within a restoration process; and \textbf{Access}, which contains fields for users with different levels of permission.

As mentioned, the smart contracts serve as our backend, featuring functions for registering vehicles, transferring ownership, adding new tasks, appending documents, querying the current vehicle information (referred to as V\textit{ehicle Card}), accessing the change history of this \textit{Vehicle Card}, granting new permissions, and more.

This also encompasses access control and privacy policies. Given the use of a single channel, we opted for Attribute-Based Access Control instead of private data collections or even multiple channels to ensure data privacy. These alternatives ensure data privacy at the organizational level rather than the end-user level. Thus, each user possesses specific attributes associated with their certificate, allowing for individual identification.

\subsection{Fabric Application and API}

To interact with the chaincode, a Fabric application was developed to support the direct invocation of the available smart contract’s functions. Fabric applications are the only place outside the network where transaction proposals can be generated, invoking specific smart contract functions. To be able to make these functions available through endpoints, we implemented this Fabric application to serve as a \gls{REST} \gls{API}, that once hosted on our server, acts as our back-end server application that exposes \gls{HTTPS} endpoints for each main function on the smart contract’s logic. These endpoints can then be invoked from other clients and applications.

This component was developed using \href{https://nodejs.org/en}{\textbf{\textit{Node.js}}} due to being the most refined option that can integrate with the \gls{SDK} provided by Hyperledger Fabric. Additionally, Node.js proves highly effective in developing web applications and microservices.

To implement the application’s \gls{API} and corresponding endpoints and functionalities we used \href{https://expressjs.com/}{\textbf{\textit{Express}}}, which is a Node.js framework, known for its simplicity and flexibility in building REST APIs. The functionalities of this API are exclusively accessible through secure data transmission via \gls{HTTPS}. This requirement ensures that all client and API communication is encrypted, enhancing data security and privacy concerns.

In terms of this application's initial configuration, it is necessary to import and integrate a connection profile, which is essentially a file describing the entire Fabric network. With this connection profile, the client is then generated for each organization, along with wallets that will store each user's certificates.

Our system's registration and authentication functions interact not only with the Fabric network but also with a user database. Thus, in addition to each user being registered on the network with the corresponding organization and obtaining a certificate for authentication, they are also registered in a \href{https://www.mongodb.com/}{\textbf{\textit{MongoDB}}} user database managed by each organization. This establishes a one-to-one relationship between entities in the Fabric network and system users. Still, it uses the expected certificates when interacting with the blockchain network.

Transitioning to interaction with the chaincode, nearly all functions available in the smart contracts have a corresponding function in our Fabric application and a corresponding endpoint in our \gls{REST} \gls{API}. Each function undergoes initial data and prerequisite validation. Then the desired function can be executed, using either the \textit{submitTransaction} function to record a new transaction on the ledger or the \textit{evaluateTransaction} for read operations that do not originate new transactions, solely querying the network.

\subsection{Media Storage}

As mentioned throughout this document, from the beginning of this project's research, we recognized that off-chain decentralized media storage would be a vital component of our implementation since blockchain technology is unsuitable for high-volume data. This component ensures that the media files of various types and formats are adequately stored and an identifier of their content, based on a cryptographic hash, is subsequently recorded in our blockchain network. This approach is consistently applied to functionalities that involve interaction with media files.

With the study's initiation, it was also promptly acknowledged that \gls{IPFS} would align with our requirements as the chosen off-chain media storage system. Hence, integrating \gls{IPFS} with our Fabric application became imperative.

\gls{IPFS} is a decentralized \gls{P2P} protocol designed for efficient and permanent storage of various file types. It ensures the uniqueness and integrity of files through the \gls{CID}, generated from a hash of their content. This low-volume \gls{CID} can be registered on our blockchain, providing a unique and immutable reference to the content on the network.

To integrate with our application, we initially explored using a public gateway; however, it quickly became apparent that this would not be a viable option. The upload and retrieval processes were time-consuming, frequent failures occurred due to high network congestion, and files were subject to garbage collection.

We then decided to employ a solution that combines the \gls{IPFS} protocol with \href{https://filecoin.io/}{\textit{Filecoin}}. Filecoin is essentially a \gls{P2P} storage service where storage providers lease their space through an agreement between parties, enabling long-term storage. We utilized the \href{https://nft.storage/}{\textit{NFT.Storage}}, which offers this service for free and is explicitly designed for the storage of \glspl{NFT}. 

To enhance file upload times, we opted to locally generate the \gls{CID} and immediately register it to the blockchain network in the background, this way ensuring minimal impact to the end-user.

\autoref{tab:oldvsnewcid} illustrates the improvements in request duration resulting from the transition to our new implementation approach of generating the \gls{CID} locally compared to the previous sequential method when simulating requests to our application through the available \gls{API} endpoints. A consistent decrease in the duration of all requests is evident, highlighting the enhanced efficiency achieved.


\begin{table}[]
\centering
\caption{Comparison of requests average duration in seconds: old approach vs new approach}
\label{tab:oldvsnewcid}
\begin{tabular}{
>{\columncolor[HTML]{EFEFEF}}p{7cm} p{2.5cm}p{2.5cm}}
\hline
\textbf{Request}                                                     & \cellcolor[HTML]{EFEFEF}\textbf{Old approach} & \cellcolor[HTML]{EFEFEF}\textbf{New approach} \\ \hline
Add new document (1MB to 5MB)                               & \hfil6.2                                                                                                        & \hfil2.9                                                                                                        \\ \hline
Add new restoration with 2 files (up to 1MB)                & \hfil6.4                                                                                                        & \hfil2.4                                                                                                        \\ \hline
Add new restoration with 5 files (up to 1MB)                & \hfil14.8                                                                                                       & \hfil2.5                                                                                                        \\ \hline
Add new restoration with large files (up to 15MB) & \hfil8.8                                                                                                        & \hfil3.1                                                                                                        \\ \hline
\end{tabular}%
\end{table}


\subsection{User Interface}

It is also important to present the details of the developed \gls{GUI} and front-end application. This is a crucial component of our solution since it is the "face" of \textit{ClassicsChain}. It provides visual functionalities for users to interact with our solution, serving classic car owners and other entities or users from restoration workshops or certification and regulation bodies. As explained previously, Angular was selected to develop this \gls{GUI}, choosing this TypeScript-based framework because it was already being used in other projects within this initiative, promoting project continuity and consistency.

\begin{figure}[htbp]
	\centering
	\includegraphics[width=1.0\linewidth]{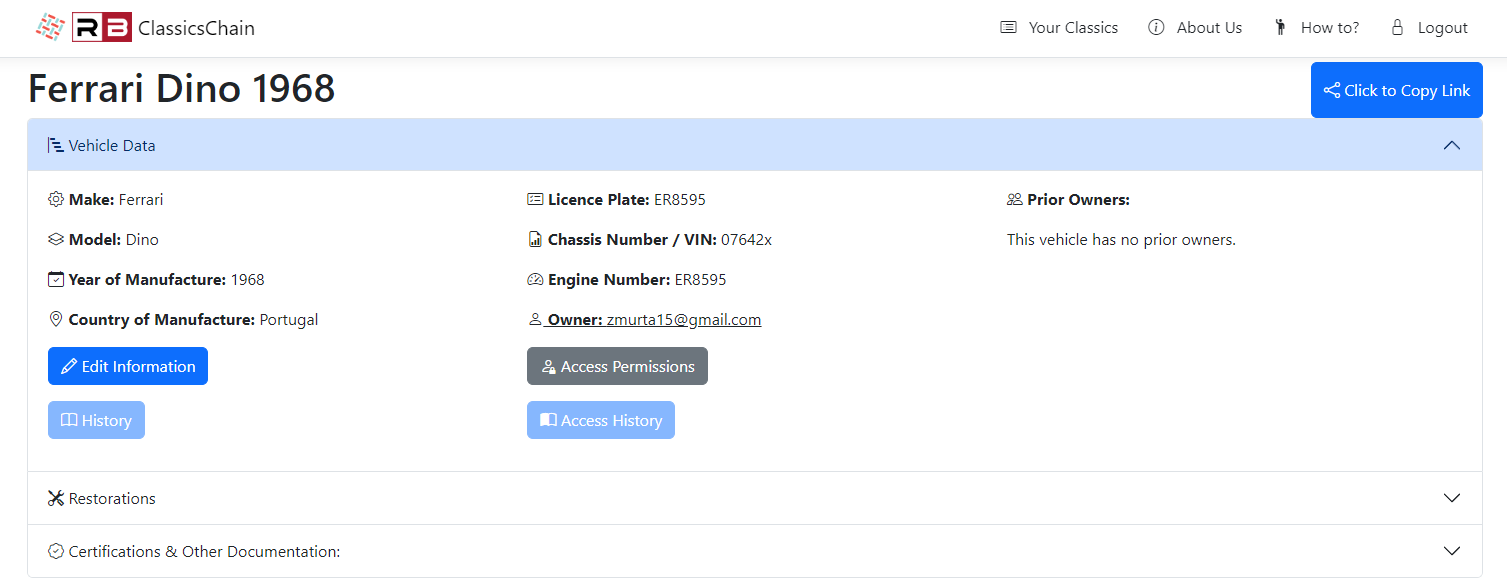}
	\caption{The "Fundamental Vehicle Data" section of the "Vehicle Card" page on the ClassicsChain website.}
	\label{fig:vehiclecard1}
\end{figure}

\begin{figure}[htbp]
	\centering
	\includegraphics[width=1.0\linewidth]{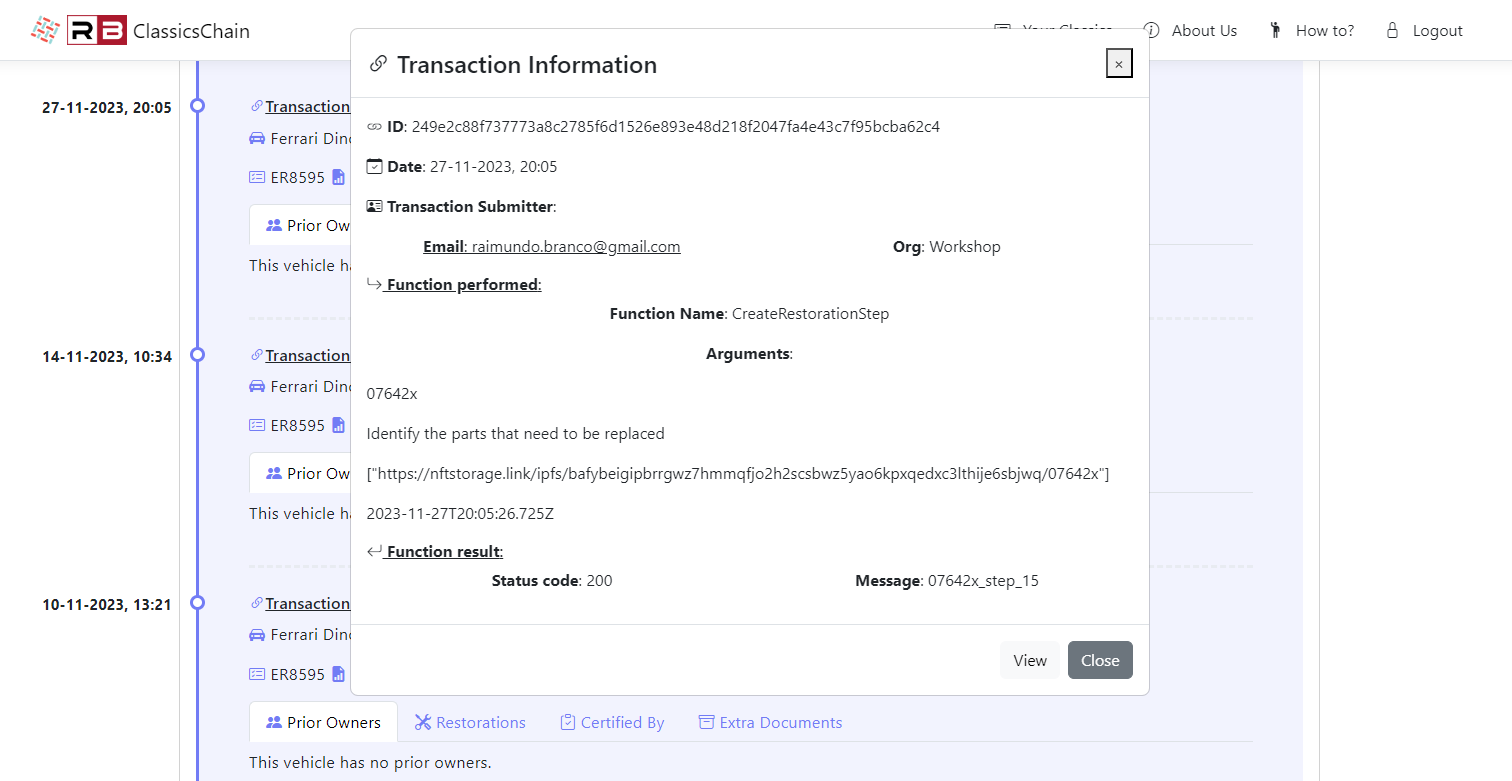}
	\caption{The "Transaction information" pop-up of the "History of the Vehicle Card" page on the ClassicsChain website.}
	\label{fig:history}
\end{figure}

For simplicity reasons, in this paper, we will only be presenting two of the most important pages of the \gls{GUI}, which are the Vehicle Card page, displayed in figure \ref{fig:vehiclecard1}, and the transaction information available for a specific transaction present on the history of the versions of the Vehicle Card, displayed in figure \ref{fig:history}.

\section{Evaluation}
\label{sec:v_and_v}

Evaluating the solution is crucial in a \gls{DSR} methodology. Therefore, it is vital to use precise and thorough methods to assess its impact, quality, and performance. 

\subsection{Performance Evaluation}

To assess the performance of our solution, our main focus centered on the \textit{ClassicsChain} Fabric application and \gls{API} through stress tests since this is the component that users will interact with. Users engage with it either through the \textit{ClassicsChain} front-end client or other clients invoking requests to our \gls{REST} \gls{API}.

Nevertheless, we also conducted evaluations of the individual performance of the Hyperledger Fabric blockchain network, our solution's cutting-edge technology. This assessment aimed to comprehend the unique performance characteristics of this component, gauge its impact on the overall system, and dissociate it from the performance of the Fabric application.

\subsubsection{Blockchain Network Evaluation}

For the individual blockchain network's performance evaluation, we employed \href{https://www.hyperledger.org/projects/caliper}{\textit{Hyperledger Caliper}} to analyze the performance through the invocation of various functions directly from the smart contracts deployed in the network.

\begin{figure}[htbp]
	\centering
	\includegraphics[width=.7\linewidth]{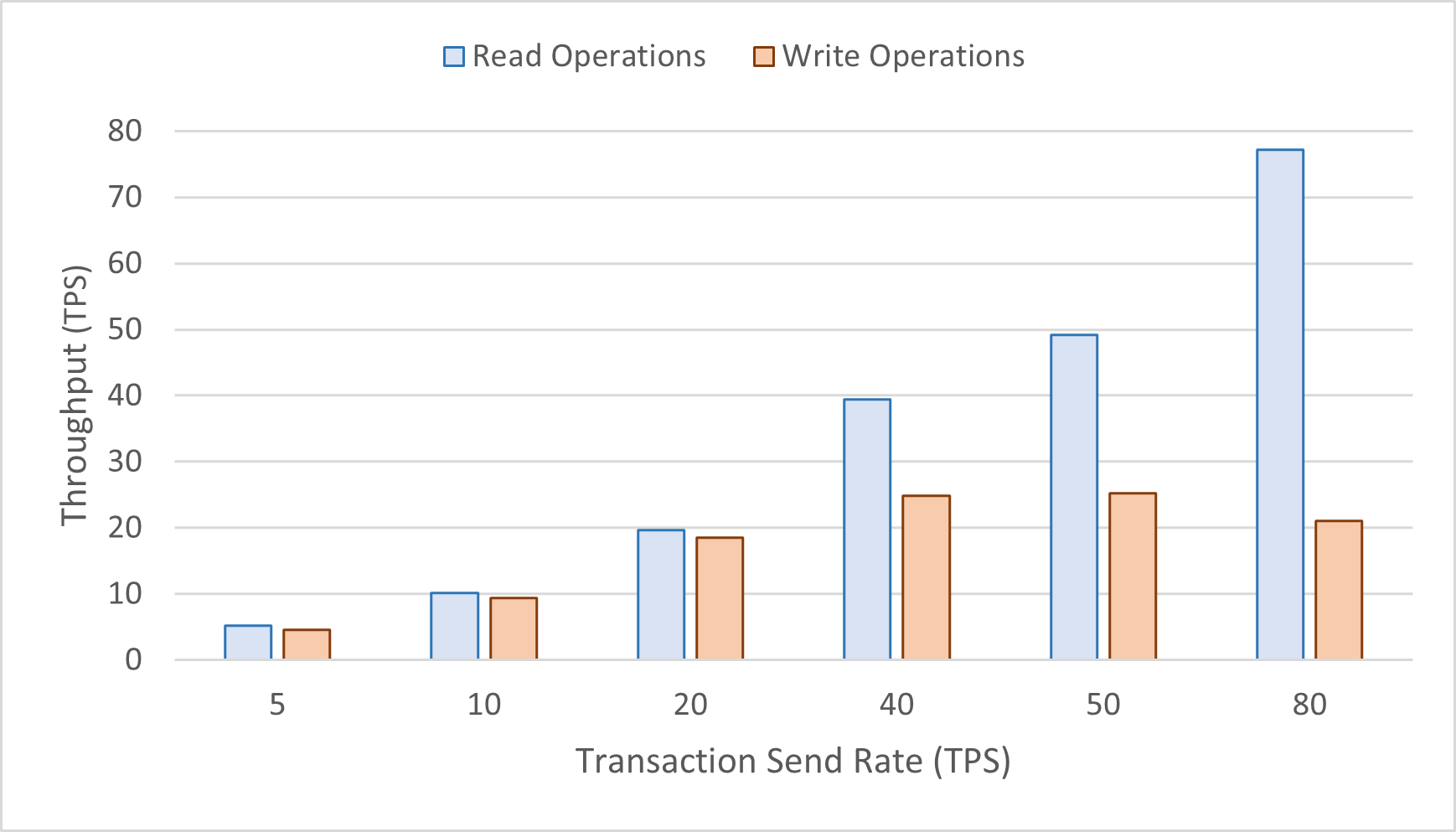}
	\caption{Throughput results of our Hyperledger Fabric blockchain network.}
	\label{fig:caliper}
\end{figure}

The results from the tests using Caliper are displayed in \autoref{fig:caliper}. These results were obtained from evaluating the most up-to-date version of the blockchain network and smart contracts, allowing us to conclude the limits of our blockchain network in terms of supported \gls{TPS} for both read and write operations, clearly defining a maximum threshold for these operations. 

We were able to determine the \gls{TPS} that our network supports, achieving a total of 70 TPS for read operations before the throughput decreases and a maximum of 22 TPS for write operations, beyond which the throughput significantly decreased and considerable failures were observed. No failures were noted in read operations.

Caliper also allows us to record latency duration in seconds. In typical network conditions and transaction send rates, we obtained average values of 0.16 seconds for read operations and 1.52 seconds for write operations.

\subsubsection{Fabric Application and API Evaluation}

Performance tests were conducted to simulate real users making requests using the Web app by employing Bash Shell scripts to communicate with the Web server via a REST API.

This evaluation aimed to determine the duration of the most important functionalities available through the respective endpoints. This assessment is essential as it evaluates the waiting time for end-users, from executing a specific request to obtaining the expected response, commonly referred to as load time.

\autoref{tab:performance} displays the minimum, average, and maximum duration times, in seconds, for the most frequently used requests, with each request being executed 30 times. Requests marked with an asterisk (*) refer to instances of that request with notable alterations, such as increased data to be uploaded or retrieved.


\begin{table}[]
\centering
\caption{Performance Evaluation: API request's duration, in seconds.}
\label{tab:performance}
\begin{tabular}{p{11cm}p{1cm}p{1cm}p{1cm}}
\hline
\rowcolor[HTML]{EFEFEF} 
\textbf{Request} & \textbf{Min.} & \textbf{Avg.} & \textbf{Max.} \\ \hline
1) Register a new user on the system & 0.81 & 0.93 & 1.28 \\
2) Login a user & 0.13 & 0.14 & 0.18 \\
3) Get the classic cars owned/authorized by a specific user (up to 30KB & 0.10 & 0.11 & 0.13 \\
3*) Get the classic cars owned/authorized by a specific user (up to 100KB) & 0.29 & 0.31 & 0.47 \\
4) Register a new classic car & 2.27 & 2.30 & 2.40 \\
5) Create a new restoration procedure (0 media files) & 2.29 & 2.32 & 2.38 \\
5*) Create a new restoration procedure (1 or 2 media files) & 2.37 & 2.42 & 2.70 \\
5**) Add a new restoration procedure (more than 5 media files) & 2.42 & 2.54 & 2.78 \\
6) Get the Vehicle Card of a specific classic (up to 50KB) & 0.09 & 0.10 & 0.10 \\
6*) Get the Vehicle Card of a specific classic (up to 100KB) & 0.25 & 0.26 & 0.30 \\
7) Get the History of the Vehicle Card of a specific classic (up to 50 KB) & 0.10 & 0.10 & 0.13 \\
7*) Get the History of the Vehicle Card of a specific classic (up to 300KB) & 0.36 & 0.43 & 0.45 \\
7**) Get the History of the Vehicle Card of a specific classic (2MB or more) & 6.27 & 8.40 & 12.14 \\
8) Grant/Revoke access permissions to a specific vehicle & 2.23 & 2.33 & 2.35 \\
9) Change a classic's owner & 2.21 & 2.46 & 2.7
\end{tabular}
\end{table}


From the obtained results, write operations tend to be more consistent in terms of duration, as the volume of exchanged data remains fixed within an expected range. This consistency holds even when involving the upload of media files, thanks to the implementation of locally generating the \gls{CID}. However, some requests that read from the blockchain take longer when the volume of data to be received increases. For instance, when requesting the version history of a Vehicle Card with around 100 versions, such a query can take more than 12 seconds.

To conclude this performance evaluation, it is discernible that the request duration times for the functionalities within our solution are acceptable and reasonable within the context of blockchain-based systems and in the context of web applications for the classic car ecosystem. These durations offer users tolerable load times, ensuring a responsive experience. Upon analyzing all available requests, the average duration was calculated, obtaining an average of 2.5 seconds for write operations and 0.75 seconds for read functions, regardless of the data size being uploaded or retrieved. These duration values fall well within the acceptable threshold, remaining under the commonly tolerated 3-second waiting time for end-users in web systems, as proposed in \cite{nah2004study}.

\subsection{Relevance and Usability Evaluation}

Regarding the evaluation of our solution's relevance and the web application's usability, we followed a methodology based on questionnaires and interviews \cite{seidman2006interviewing}.

The participants in this evaluation included: one classic car owner; an automotive industry manager with extensive experience; the director of ACP Clássicos, a division within \gls{ACP} that certifies vehicles of historical interest; and both a business sales manager and commercial director from BASF, focusing on the BASF's Glasurit brand specialized in providing high-quality automotive coatings, used for refinishing classic car vehicles.

Initially, we presented a questionnaire on \textit{ClassicsChain} functionalities without revealing the ongoing solution. Responses supported the need for our work, indicating potential enhancements for existing classic automotive solutions.

Then, we explained how \textit{ClassicsChain} functions and demonstrated the majority of the features available in the web app. Furthermore, participants were given the chance to interact with the platform actively.

Following that, we used another questionnaire to evaluate the \gls{SUS} \cite{susArticle}. This questionnaire assessed the usability and user experience of the \textit{ClassicsChain} \gls{GUI}. The results of the questions can be consolidated into a total score. As each of the ten questions has 5 possible answers, these can be scored ranging from 0.0 to 10.0 with a 2.5 interval, resulting in a possible score for each question of 0.0, 2.5, 5.0, 7.5, or 10.0 depending on the tone of the question being positive or negative which alters the order of this sequence. By summing the values of all questions, a final score ranging from a minimum of 0.0 to a maximum of 100.0 is calculated. This cumulative value can then be better converted into a grade, ranging from A+ to F, or represented as a percentile \cite{susArticle}.

\begin{figure}[htbp]
	\centering
	\includegraphics[width=.7\linewidth]{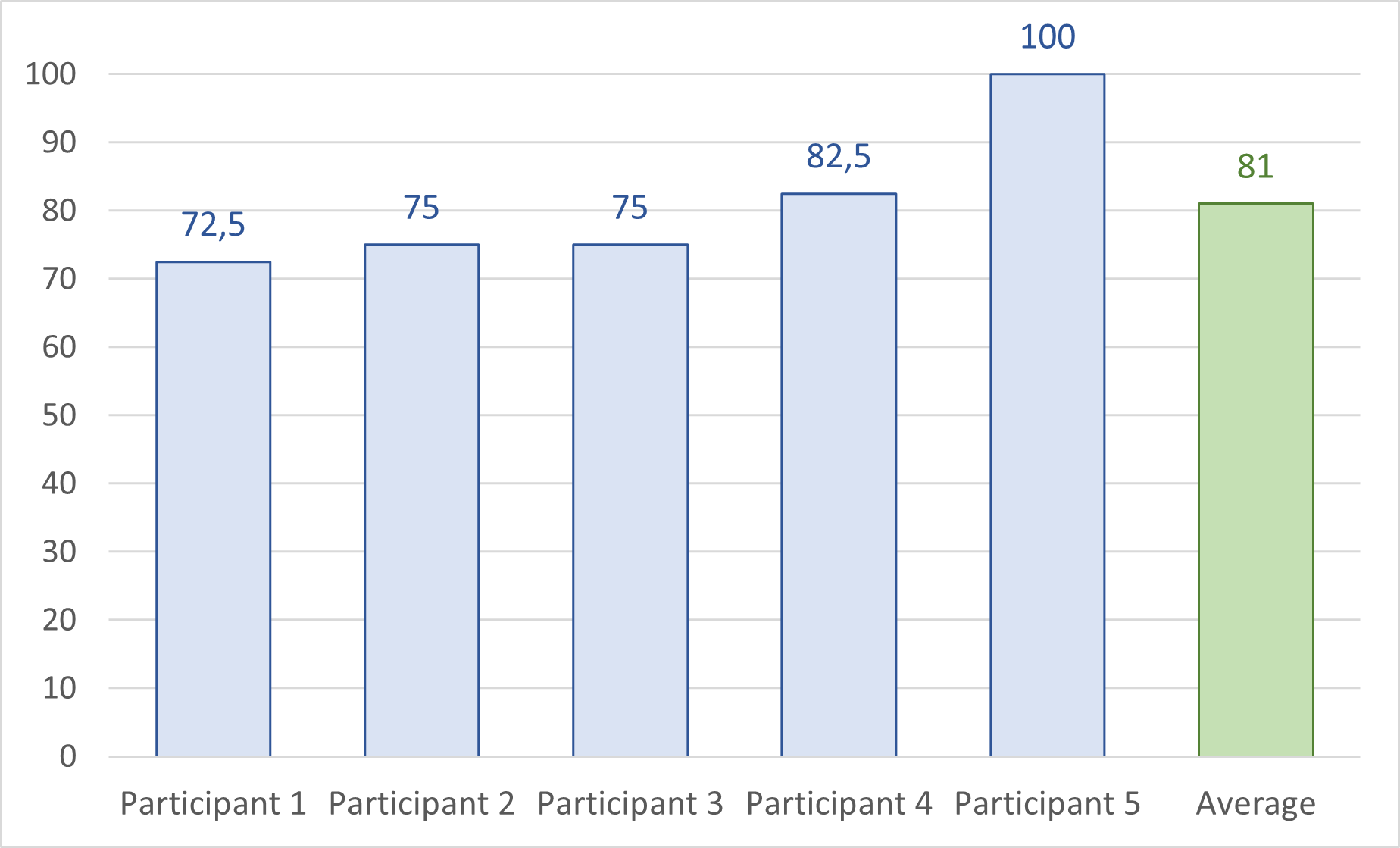}
	\caption{\gls{SUS} results for the \textit{ClassicsChain} web platform.}
	\label{fig:susResults}
\end{figure}

The participants' final scores in the \gls{SUS} section of the questionnaire are presented in \autoref{fig:susResults}. The average score yields a final result of 81, corresponding to a grade of A when converted. This score is considered above average for user experience and usability, surpassing the score of 80, as suggested by J. R. Lewis \cite{susArticle}.

Additionally, we presented another questionnaire to gather opinions about the overall concept of the platform. The answers to these questions further emphasize the need for the work undertaken in this research and the relevance of \textit{ClassicsChain} and its distinctive features. Particularly noteworthy are the positive responses indicating participants' intentions to utilize the platform for future restoration procedures and their willingness to recommend it to other enthusiasts and entities involved in the classic automobile context.

\subsubsection{Threats to the Evaluation}

Upon completion of the relevance and usability assessment, it is important to clarify the validation threats identified. Moderator presence in interviews and questionnaires poses a validation threat, potentially biasing participant responses to please or displease the moderator and impact the research.

Moreover, due to time and availability constraints, the participant sample was restricted to five individuals, potentially affecting result precision.

Additionally, despite being unrelated to our work or the solution developed, some participants had prior exposure to the workshop where the solution was developed, which might introduce a positive bias in responses to maintain a favorable relationship.

Some participants could not explore the platform themselves. While all participants observed the platform in operation, mere observation might not suffice to evaluate certain usability aspects comprehensively. The absence of hands-on experience could have influenced the outcomes, potentially presenting outcomes that are either more favorable or less accurate. Additionally, participants used the interface in a controlled interview environment, impacting result authenticity compared to their natural interaction.
\section{Conclusions} \label{sec:conclusions}

\subsection{Conclusions} \label{sub:conclusions}

The goal of the reported research is to develop and validate a solution for capturing and preserving key information about classic cars, particularly for their restoration processes, with an emphasis on authenticity. 

The proposed solution prioritizes trustworthy, immutable, and transparent information. It aims to facilitate information sharing, foster industry collaboration and promote best practices. The work, carried out using a \gls{DSR} methodology, included a multivocal literature review, the exploration of cutting-edge technologies, and evaluation procedures along several perspectives (performance, relevance and usability).

The solution developed, \textit{ClassicsChain}, is a blockchain-based application that records technical specifications, restoration evidence, certificates, ownership details, and other critical documentation for classic cars. Powered by a Hyperledger Fabric blockchain network, \textit{ClassicsChain} ensures immutability, transparency, and unique features, including a chronological history, access control, and integration with off-chain media storage. For tracking the restoration progress, it was integrated with the \textit{Charter of Turin Monitor}, a BPMN process-based platform developed within our research team \cite{PedroMoura2023}.

By bringing the benefits of blockchain-based systems into the classic car ecosystem, our proposal catalyzes its digital transformation because preserving authenticity is crucial for fostering classic cars' historical, cultural, and collector value, honoring their original design, craftsmanship, and engineering excellence.

\subsection{Future Work} \label{sub:futureWork}

The implementation of a mediated entity verification mechanism would increase the credibility of our solution and thus of the recorded information. In this mechanism, users would have to confirm their identity with some form of unique identification document, similar to the processes in banking or government applications.

It would also be interesting adding information on events where classic cars often participate such as \href{https://en.wikipedia.org/wiki/Concours_d%27Elegance}{Concours d'Elegance} or competitions, as well as information about parts added, removed, or modified in a particular restoration process.

Another feature that is being considered based on feedback collected during the evaluation procedures is the implementation of an alert mechanism when data is subsequently modified. This would facilitate the fact-checking process.

Another possible consideration would be to expand the range of features for users, such as implementing a chat for communication among them (e.g. between car owners and potential buyers or workshops).

It would also be relevant to evaluate the web application with a larger sample of users.

Last but not least, it would make sense to transition our blockchain network infrastructure, as well as the \gls{IPFS} network, to an infrastructure with nodes that are fully managed by the represented organizations. This shift would give the respective organizations or entities full control and management of these nodes, as opposed to relying on the services offered by a cloud service provider. However, its feasibility may be questionable, as it is a complex task requiring advanced expertise and dedicated resources that may not be available in these organizations.

\begin{acks}
This work was produced with the support of INCD funded by FCT and FEDER under the project 01/SAICT/2016 nº 022153, and partially supported by NOVA LINCS (FCT UIDB/04516/2020), and ISTAR-Iscte (FCT UIDB/04466/2020 and UIDP/04466/2020).
\end{acks}

\bibliographystyle{ACM-Reference-Format}
\bibliography{main}

\end{document}